\documentclass[12pt]{article}

\usepackage{amscd}

\usepackage{verbatim}

\usepackage{amssymb}

\usepackage{amsmath}

\newcommand{\be}{\begin{equation}}
\newcommand{\ee}{\end{equation}}
\newcommand{\ben}{\begin{eqnarray}}
\newcommand{\een}{\end{eqnarray}}
\newcommand{\ba}{\begin{eqnarray}}
\newcommand{\ea}{\end{eqnarray}}

\newcommand{\bi}{\begin{itemize}}
\newcommand{\ei}{\end{itemize}}

\newcommand{\lb}{\left (}
\newcommand{\rb}{\right )}

\newcommand{\p}{\partial}

\begin{document}

\begin{center}

\vspace{24pt} { \large \bf Investigating stability of a
class of black hole spacetimes under Ricci flow } \\

\vspace{30pt}

\vspace{30pt}

{\bf Suvankar Dutta$^{(a)}$}\footnote{pysd@swan.ac.uk}, {\bf V
Suneeta$^{(b)}$}\footnote{vardarajan@math.ualberta.ca}

\vspace{24pt} 
{\footnotesize $^{(a)}$Dept of Physics, Swansea University, Swansea,
UK.}\\ \vspace{14pt} 
{\footnotesize $^{(b)}$Dept of Mathematical and Statistical Sciences \\ and\\
The Applied Mathematics Institute,\\ University of Alberta,
Edmonton, AB, Canada T6G 2G1.}

\end{center}
\date{\today}
\bigskip

\begin{center}
{\bf Abstract}
\end{center}

\noindent We prove the linear stability of Schwarzschild-Tangherlini
spacetimes and their Anti-de Sitter counterparts under Ricci flow
for a special class of perturbations. This is useful in the choice
of suitable initial conditions in numerical Ricci-flow-based
algorithms for obtaining new solutions to the Einstein equation when
the cosmological constant is zero or negative. The Ricci flow is a
first-order renormalization group (RG) flow in string theory, and
its solutions are believed to approximate string field theory
processes in certain cases. Thus this result offers insights into
the off-shell stability of these Euclidean black hole geometries in
string theory, as well as in the Euclidean path integral approach to
quantum gravity.

\newpage

\section{Introduction}
\setcounter{equation}{0}

There have been numerous studies of the classical stability of black
hole and black brane spacetimes under gravitational perturbations.
It is well-known that Schwarzschild black holes are classically
stable \cite{sbhstable}. The classical stability of the higher
dimensional Schwarzschild-Tangherlini and
AdS-Schwarzschild-Tangherlini black holes for a class of
perturbations has been shown by Gibbons and Hartnoll \cite{GH} and a
stability result under all classes of perturbations obtained by
Ishibashi and Kodama \cite{ishibashi}.

There are also many distinct motivations for studying {\em
off-shell} stability of black hole spacetimes. What is meant by the
term `off-shell' varies depending on the context. The subject of
this paper is a study of the off-shell stability (in a certain
sense) of Schwarzschild-Tangherlini and
AdS-Schwarzschild-Tangherlini black holes. More precisely, we study
the linear stability of these black hole spacetimes under Ricci
flow. We discuss the various motivations for this study, and in the
process, we review the various notions of off-shell stability and
their connections to physics. \vskip 0.3cm {\bf Euclidean path
integral formulation of quantum gravity:}\vskip 0.3cm

One of the earliest motivations comes from the Euclidean path
integral formulation of quantum gravity. Classical configurations
are stationary points of the gravity action, and off-shell
configurations mediate in a quantum tunnelling from one classical
configuration to another. In this context, off-shell perturbations
of a classical configuration that make the action negative lead to
instabilities at least in the semiclassical approximation to quantum
gravity. This has motivated a study of the off-shell linear
(in)stability of the Euclidean Schwarzschild instanton --- this
instanton was shown by Gross, Perry and Yaffe (GPY) \cite{GPY} to
have an unstable off-shell mode. Computationally, GPY find a
normalizable eigenmode of the Lichnerowicz Laplacian (\ref{2.6}) for
the Euclidean Schwarzschild metric with a negative eigenvalue. The
perturbation described by this negative mode is off-shell (the
perturbed geometry is not a solution to the Einstein equation in
this linearized approximation). This perturbation is therefore
evidence of the off-shell instability of the Euclidean Schwarzschild
instanton in semiclassical gravity. It was argued by Reall
\cite{reall} that this `quantum' instability is in fact the same as
the classical Gregory-Laflamme instability of the uncharged black
$p$-brane whose metric splits into the four dimensional
Schwarzschild metric and the flat metric in $p$ dimensions
\cite{GL}. Gregory and Laflamme assume a special ansatz for the
perturbation (periodic along the $p$ directions) such that the
classical unstable mode of the $p$-brane is the same as the negative
eigenmode of the Lichnerowicz Laplacian on the four dimensional
Schwarzschild instanton obtained after compactifying $p$ directions.
Arguments about the existence of a GPY-type negative mode for the
Lichnerowicz Laplacian of higher dimensional black holes can be
found in \cite{prestidge}. These results have close connections to
the linear stability of these black holes under Ricci flow --- the
operator describing the linearized flow of the perturbation of the
black hole spacetime is the Lichnerowicz Laplacian (or related to
it). Thus instability under Ricci flow can be due to the
Lichnerowicz operator having a negative mode. This would then also
have implications for the classical stability of corresponding $p$
branes.\vskip 0.3cm

{\bf Numerical Ricci flow-based algorithm for obtaining new
solutions to the Einstein equation:} \vskip 0.3cm

The Ricci-de Turck flow ((\ref{2.1}), with $\alpha' = 2$) is a
geometric flow equation well-studied in mathematics and was used to
successfully resolve the Poincare conjecture. Ricci-flat metrics are
fixed points of this flow, which is a nonlinear parabolic PDE.
Recently, a numerical algorithm has been proposed for obtaining new
solutions to the vacuum Einstein equation --- this aims to use the
Ricci flow through static spacetimes to converge to new Ricci-flat
solutions of the Einstein equation \cite{hw}, \cite{hwk}. The
algorithm involves a numerical simulation of Ricci flow on static
spacetimes starting from fine-tuned initial data --- the hope being
that if the initial data are appropriately chosen, the Ricci flow
will converge to a Ricci flat fixed point. Without careful choice of
initial data, the Ricci flow could become singular, or may not
converge to a Ricci-flat fixed point. This problem is illustrated by
a numerical simulation by Headrick and Wiseman \cite{hw} of Ricci
flow of a (Euclidean) Schwarzschild black hole in a radial box. If
the ratio of horizon radius to box radius is less than $2/3$, this
black hole is the `small' black hole and its Lichnerowicz operator
has a negative mode (as discussed before). If this ratio is greater
than $2/3$ (the large black hole), this negative mode is not
present. The negative mode of the small black hole is not an
artefact of putting the black hole in a box. When the box radius is
taken to infinity, this mode persists and is precisely the GPY mode.
In \cite{hw}, the small black hole perturbed by this unstable mode
is chosen to be the initial data for a numerical Ricci flow
simulation. Surprisingly the authors find that the numerical
simulation yields different results depending on whether the
perturbation is added to the small black hole metric, or subtracted
from it. In one case, the metric flows to the large black hole and
in the other case, it becomes singular. Thus the moral of this
example seems to be that for the algorithm to work, the initial
metric cannot be an arbitrary static metric. For example, one could
choose a perturbation of a known Ricci-flat metric, which leads to
an instability, causing a flow away from the known Ricci-flat
metric. Thus knowledge of both the {\em number} of unstable modes of
Ricci-flat static metrics and the exact {\em form} of the unstable
mode are required for choosing the right initial data that will lead
to convergence of the Ricci flow. \footnote{This problem is not
faced in a similar simulation of K\"ahler-Ricci flow on the third
del-Pezzo surface as in \cite{doran}. In this case, a theorem of
Tian and Zhu \cite{tian} guarantees that starting from any initial
K\"ahler metric on this surface (obeying a certain condition), the
flow converges exponentially to a K\"ahler-Einstein metric.} This
instability under Ricci flow is an `off-shell' instability in a
sense, since the perturbed metric does not solve the Einstein
equation. The algorithm in \cite{hwk} could be extended to finding
new solutions to Einstein equation with cosmological constant by
using the flow (\ref{2.3}) instead of Ricci flow. This is one of the
main motivations for this paper, where we examine the linear
stability of Schwarzschild-Tangherlini black holes (and their AdS
analogues) under Ricci flow or the flow (\ref{2.3}). As we also show
in section IV, this is related to the problem of eigenmodes of the
Lichnerowicz Laplacian on these spacetimes. \vskip 0.3cm

{\bf Off-shell stability in string theory:} \vskip 0.3cm Finally we
discuss yet another unrelated motivation for studying the stability
of the Schwarzschild-Tangherlini black holes under Ricci flow. Ricci
flow arises naturally as a first-order world-sheet renormalization
group (RG) flow in closed string theory. It has been conjectured
that off-shell processes in string field theory such as tachyon
condensation are approximated qualitatively by solutions to
world-sheet RG flows in string theory (see \cite{HMT} for a review).
There is some evidence for this --- for example, the tachyon
condensation process leading to the geometry change $C/Z_n
\rightarrow C$ is described in \cite{APS}, and the exact solution to
Ricci (first-order RG) flow which describes this geometry change is
one of the K\"ahler-Ricci solitons of H-D. Cao \cite{cao} (this
solution also appears in different coordinates in \cite{GHMS}). One
of the arguments in support of this conjecture is that in many
cases, in the CFT describing the fixed point of the RG flow (a
vacuum or on-shell geometry), it is possible to construct operators
that are relevant perturbations of the fixed point (causing an RG
flow) and as well, tachyonic. This suggests an investigation of the
off-shell (in)stability of a vacuum geometry by studying its
(in)stability under a suitable RG flow --- for example, the Ricci
flow. This proposal was suggested in \cite{suneeta} where the linear
stability of Euclidean AdS space (Hyperbolic space $\mathbf H^n$)
under Ricci flow was proved. We continue this program here by
investigating stability of black hole geometries under Ricci flow.
The Euclidean Schwarzschild-Tangherlini geometry is a fixed point of
the first-order world-sheet RG (Ricci) flow. Of course, the
Lorentzian black hole geometry has a curvature singularity, and
therefore, the perturbative sigma model anaysis of this target space
geometry ought to break down at the singularity. However, it is
useful to recall that the standard $\beta$ function and RG flow
computations for world-sheet sigma models assume that the metric is
Riemannian (Wick-rotated Lorentzian metric). In fact, the
first-order RG flow is not well-behaved when considered as a PDE on
general Lorentzian spacetimes. Thus the RG flow should be thought of
as a flow through a class of spacetimes for which Wick-rotation
makes sense (static or at least stationary) --- such as the
spacetime outside the horizon for the Schwarzschild-Tangherlini
black holes. \footnote{ Since the first-order (Ricci) flow preserves
isometries, if the initial spacetime is static, we are guaranteed
that the solution will be in this class.} Analogous to classical
stability calculations, it would be interesting to analyze stability
of the Schwarzschild metric treating the horizon as a boundary. If
the metric is unstable under first-order RG flow, then the natural
question to pursue is what the end-point of the flow is.

\vskip 0.3cm In this paper, we prove the linear stability of (the
spacetime outside the horizon of) the Schwarzschild - Tangherlini
black holes and their AdS counterparts under Ricci flow for a
special class of static perturbations. We have already seen several
motivations for this stability investigation. We also show that
given an arbitrary static perturbation, these black holes do not
have any instability in the flow of the trace of the perturbation.
In section II, we discuss the Ricci flow and a linearized analysis
of the evolution of a perturbation of a Ricci-flat geometry under
Ricci flow. We also describe how the evolution of perturbations of
any Einstein metric under Ricci flow can be studied. We then discuss
details such as gauge-fixing for simplifying the analysis of the
flow of the perturbation. In section III, we restrict to a special
class of perturbations and discuss the flow equation for the
perturbation of Schwarzschild-Tangherlini black holes (and their AdS
counterparts) in this case. Section IV contains the analysis of the
flow equation by assuming an ansatz, and proving that there are no
normalizable unstable perturbations of this form for both classes of
black holes. Sections III and IV are the analogues of the classical
stability results of Gibbons and Hartnoll \cite{GH}, and we try to
follow similar notation wherever possible. Section V provides a more
rigorous argument (without the need to choose a special ansatz) that
shows that perturbations of compact support do not grow under the
linearized flow. This is analogous to the result of Wald \cite{wald}
in the proof of the classical stability of the Schwarzschild black
hole. Finally, in the last section, we summarize our results and
discuss how to extend them to investigate stability of the
Schwarzschild-Tangherlini black holes under all classes of
perturbations.

\section{A linearized analysis of stability}

 \setcounter{equation}{0}

The Ricci flow, which is a subject of active research in
mathematics, is also the simplest lowest-order (in square of string
length $\alpha'$) RG flow of the world-sheet sigma model for closed
strings. In this context, the Ricci flow is the flow of the metric
of the target space with respect to the RG flow parameter $\tilde
\tau$. As mentioned earlier, this flow is not well-behaved when
considered as a flow through general Lorentzian spacetimes. It
should be thought of as a flow through Riemannian geometries which
are Wick-rotated Lorentzian spacetimes. Both in physics and
mathematics, we are interested in a flow of metrics mod
diffeomorphisms.\footnote {In the mathematics literature, $\alpha' =
2$.} Therefore all flows related to each other by $\tilde \tau$
dependent diffeomorphisms generated by a vector field $V$ are
equivalent; we write the generic flow in this class
--- the {\em Ricci-de Turck} flow as:
\begin{eqnarray}
\frac{\partial \tilde g_{ab}}{\partial \tilde \tau} = - \alpha'
(\tilde R_{ab} + \tilde \nabla_{a} V_b + \tilde \nabla_{b} V_a )
\label{2.1}
\end{eqnarray}
Ricci flow in two and three dimensions on compact manifolds is now
well-understood
--- with curvature conditions on an initial geometry, much is known
about the limiting geometry under Ricci flow \cite{chow},
\cite{topping}. This is not the case for solutions to Ricci flow in
higher dimensions, or on noncompact manifolds. Stability results for
geometries under small perturbations are therefore useful in such
cases (for some stability results for geometries under Ricci flow,
see \cite{rfstability}, \cite{suneeta} and references therein).

The background metrics whose linear stability problem we are
interested in are a class of static metrics on $D$ dimensional
spacetimes; given by
\begin{equation}
ds^2 = - f(r) dt^2 + \frac{dr^2}{f(r)} + r^2 d\tilde s_{d}^2
\label{2.2}
\end{equation}
where $d\tilde s_{d}^2$ is a Riemannian metric on a $d = D-2$
dimensional `base' manifold $B$. Further these metrics are Einstein,
and are thus solutions to the Einstein equations with a cosmological
constant.

The flow of Lorentzian spacetimes  is not a well-posed problem in
general even in a linearized approximation. We will restrict
ourselves to the class of static perturbations of the spacetimes ---
for this class, the (linearized) flow is well-posed.

It is more convenient to study stability of a geometry under a
geometric flow when it is a fixed point of the flow. However
Einstein metrics with non-zero Ricci tensor (i.e., non-zero
cosmological constant) are not fixed points of Ricci flow, so we
describe a technique which can be used to study stability of such
metrics. This is useful in extending the algorithms in \cite{hwk} to
obtain new Einstein spacetimes with non-zero cosmological constant.
Further, it is expected that at least some Einstein metrics may be
fixed points of string theory RG flows with background fields
($AdS_3$ is a fixed point of RG flow with a $B$ field). In the
absence of a full knowledge of string theory $\beta$ functions with
background fields (like the RR field), we can still discuss a notion
of stability of Einstein metrics under Ricci flow (i.e., is the
Einstein metric an {\em attractor} on the space of solutions to
Ricci flow?). The result is expected to be indicative of stability
under an appropriate RG flow at least with respect to metric
perturbations.

Einstein metrics have a simple evolution under Ricci flow --- the
metric either expands or contracts uniformly (depending on the sign
of the cosmological constant) by a conformal factor. One can study
if perturbations of the Einstein space decay, and if the perturbed
geometry approaches the Einstein space under Ricci flow (up to
overall scale). This notion of stability is called {\em geometric
stability} of the Einstein space with respect to Ricci flow. Let the
Einstein metric have Ricci tensor $R_{ab} = c (d+1) g_{ab}$.

Given the Ricci-de Turck flow (\ref{2.1}), we consider the flow
\begin{eqnarray}
\frac{\partial g_{ab}}{\partial \tau } = - \alpha' [ R_{ab} - c
(d+1) g_{ab} ] \label{2.3}
\end{eqnarray}
whose solutions are related to those of (\ref{2.1}) by
\begin{eqnarray}
\tilde \tau = \frac{1}{\alpha' c}~~ e^{\alpha' c (d+1) \tau}, \nonumber \\
\tilde g_{ab} = e^{\alpha' c(d+1) \tau} g_{ab}. \label{2.4}
\end{eqnarray}

The flow (\ref{2.3}) has the Einstein metrics with $R_{ab} = c (d+1)
g_{ab}$ as its fixed points. First we study the linear stability of
the Einstein space under this flow. Then, by the rescalings
(\ref{2.4}) this leads to a result on geometric stability of the
space under the Ricci flow (\ref{2.1}). Ricci-flow-based numerical
algorithms such as \cite{hwk} to obtain solutions to the vacuum
Einstein equations can be easily generalized to the flow (\ref{2.3})
in order to obtain new solutions to the Einstein equation with
non-zero cosmological constant.

The linear stability problem under the flow (\ref{2.3}) can be set
up as follows: Given a background metric $g_{ab}$, a perturbed
metric $g_{ab}^{p} = g_{ab} + h_{ab}$. One studies the evolution of
the perturbed metric $g^{p}$ in a linearized approximation. The
linearized flow for the perturbation $h$ is
\begin{eqnarray}
\frac{\partial h_{ab}}{\partial \tau} = \frac{\alpha'}{2} [ -
(\Delta_{L}  h)_{ab} + \nabla_{a}\nabla_{b} H -  \nabla_{a}(
\nabla^{c}h_{cb}) - \nabla_{b}( \nabla^{c}h_{ca}) \nonumber \\ + 2
\nabla_{a} V_b + 2  \nabla_{b} V_a  + 2 c (d+1) h_{ab}].~~~
\label{2.5}
\end{eqnarray}
Here and in what follows, all covariant derivatives are taken with
respect to the background metric $g$. $H = g^{ab} h_{ab}$ is the
trace of the perturbation.
\begin{eqnarray}
(\Delta_{L} h)_{ab} = - \Delta h_{ab} + 2 R_{~abc}^{d} h_{~d}^{c} +
 R_{a}^{c} h_{bc} +  R_{b}^{c} h_{ac} \label{2.6}
\end{eqnarray}
is the Lichnerowicz laplacian acting on symmetric 2-tensors (all
curvature tensors being those of the background metric).The
convention we follow for the Lichnerowicz laplacian is that of the
physics literature, and differs from the mathematics one by a
negative sign.

We can choose $V$ so that we get rid of the divergence terms in
(\ref{2.5}), namely $\frac{\alpha'}{2} \left ( \nabla_{a}
\nabla_{b}H - \nabla_{a}(\nabla^{c}h_{bc}) -
\nabla_{b}(\nabla^{c}h_{ca}) \right )$. This leads to the simplified
flow
\begin{eqnarray}
\frac{\partial h_{ab}}{\partial \tau} = -
\frac{\alpha'}{2}[(\Delta_{L} h)_{ab} - 2 c (d+1) h_{ab}] = (\mathbf
Lh)_{ab}; \label{2.7}
\end{eqnarray}

$\mathbf L = - \frac{\alpha'}{2}[(\Delta_{L} - 2 c (d+1)]$.

\vskip 0.3cm We now need to `fix gauge' in order to simplify our
problem. The following results are useful for gauge-fixing in
problems involving linearized stability analysis of Einstein spaces:
\vskip 0.3cm

1. Let $h_{ab}$ be a trace-free perturbation, i.e., $g^{ab}h_{ab} =
0$. $g^{ab}(\Delta_{L} h)_{ab} = - \Delta H = 0$, and therefore
$(\mathbf L h)_{ab}$ is also trace-free.\\

2. If the background metric $g$ is Einstein, and $\nabla^{a} h_{ab}
= 0 $ (i.e., the perturbation is transverse), then $\nabla^a
(\Delta_L h)_{ab} = 0 $ and
therefore, $\nabla^{a} (\mathbf Lh)_{ab} = 0$. \\
{\em Proof :} Consider $\nabla^a (\Delta_L h)_{ab}$. Let $h_{ab}$ be
a transverse perturbation.
\begin{eqnarray}
\nabla^{a} (\Delta_L h)_{ab} &=& 2
\nabla^{a}(R_{~abd}^{c})h_{~c}^{d} + 2 R_{~abd}^{c}
\nabla^{a}(h_{~c}^{d}) \nonumber \\ &+& \nabla^{a}(R_{ca})
h_{~b}^{c} + R_{ca} \nabla^{a}h_{~b}^{c} +
\nabla^{a}(R_{cb})h_{~a}^{c} - \nabla^{a }(\Delta h)_{ab}.~~~~
\label{2.8}
\end{eqnarray}
Then,
\begin{eqnarray}
\nabla^{a }(\nabla^{c} \nabla_{c} h)_{ab} &=& g^{cd}g^{ae} (\nabla_d
\nabla_e \nabla_c )h_{ab} \nonumber \\&-& R^{ad} \nabla_d h_{ab} +
R^{dc} \nabla_{c} h_{db} - R_{~bec}^{d} \nabla^c
h_{~d}^{e};\nonumber \\
&=& \nabla_{c} (R^{dc})h_{db} - \nabla^{c}(R_{dbac}) h^{ad} -
R_{dbac} \nabla^{c}(h_{ad}) \nonumber \\ &+& R^{dc} \nabla_{c}
h_{db} - R_{dbac} \nabla^{c} h^{ad}. \label{2.9}
\end{eqnarray}
Now using the Bianchi identity $\nabla_{a} R_{~cbd}^{a} =
\nabla_{d}R_{cb} - \nabla_{b} R_{cd}$, and (\ref{2.8}) and
(\ref{2.9}), we obtain that
\begin{eqnarray}
\nabla^{a} (\Delta_L h)_{ab} &=& (\nabla_{d} R_{cb} + \nabla_{c}
R_{db} - \nabla_{b} R_{cd}) h^{dc}. \label{2.10}\end{eqnarray}

This means that when the background metric $g$ is Einstein,
$\nabla^{a} (\Delta_L h)_{ab} = 0$ (in fact, this is true whenever
the right hand side of (\ref{2.10}) is zero even if $g$ is not
Einstein). It obviously follows that $\nabla^{a} (\mathbf Lh)_{ab} =
0$.  It therefore follows that if $h_{ab}$ is transverse and
traceless (TT), then so is $(\mathbf Lh)_{ab}$. \\

3. Consider a perturbation of the form $h_{ab} = \nabla_a W_b +
\nabla_b W_a$, a `pure divergence'. Inserting this in the
right-hand-side of (\ref{2.5}) with $V=0$, when the background
metric is Einstein, we get zero. Pure divergence perturbations do
not flow; they are `zero modes'of the linearized flow. Thus there is
no loss of generality in taking the perturbation $h$ to be
transverse. This is similar to gauge-fixing in relativity, but the
difference is that the trace of the perturbation cannot be gauged
away in Ricci flow stability problems.\\

A choice of gauge where the perturbation is transverse was used in
the proof of linear stability of Euclidean AdS space
($\mathbf{H^n}$) under (\ref{2.3}) (and consequently under Ricci
flow) in \cite{suneeta}. We briefly sketch some steps in this
computation that are similar to section V of our paper. The strategy
is to define `energy integrals' on $M = \mathbf{H^n}$, given by \be
E^{(K)} = \int_M |(\mathbf{L^{K}}h)_{ab}|^{2} ~dV, \label{2.11} \ee
and prove an upper bound on these integrals under the flow (in terms
of their initial values). Here the notation $(\mathbf{L^{2}}h)_{ab}$
indicates, for example $(\mathbf{L L}h)_{ab}$. In fact, it is
possible to prove these integrals go to zero as $\tau \rightarrow
\infty$. This is then used to prove that a certain {\em Sobolev
norm} of the perturbation goes to zero as well in this limit, where
the Sobolev norm $\parallel h \parallel_{k,2}$ is defined by
\begin{eqnarray}
\left ( \parallel h \parallel_{k,2} \right )^2 &=& \int_M |h_{ij}|^2
~dV + \int_M |\nabla_{p_1} h_{ij}|^2 ~dV + ....\nonumber \\&&...+
\int_M |\nabla_{p_1}....\nabla_{p_k} h_{ij} |^2 ~dV. \label{2.12}
\end{eqnarray}
$|h_{ij}|^2$, for example, is the square of the (pointwise) tensor
norm of the perturbation, i.e., $h^{ij}h_{ij}$. The Sobolev norm
goes to zero under the flow for all $k$. The final step in the proof
uses a {\em Sobolev inequality} on $\mathbf{H^n}$ that implies that
when this Sobolev norm goes to zero, the perturbation and all its
derivatives go to zero {\em pointwise} in $M$.\footnote{ Note that
even if the Sobolev norm goes to zero, $h$ or its derivatives could
still be non-zero on a set of measure zero, therefore we need a
Sobolev inequality to argue that they go to zero pointwise in $M$.}
This program is hard to implement to prove stability of other
Einstein metrics partly because the bound on the energy integrals
(\ref{2.11}) was only possible due to the simple form of Riemann
curvature for $\mathbf{H^n}$. Also, Sobolev inequalities are not
known for most other Einstein manifolds. Nevertheless, the analysis
in section V of our paper bears some similarities to the steps
above, where we derive and use a very simple Sobolev inequality.

Another strategy is to split a general perturbation into a
trace-free part and a part proportional to the trace; $h_{ab} = {\bf
H}_{ab} + \frac{H}{d+2} g_{ab}~(~{\bf H}_{ab}$ denotes the
trace-free part). Then the flow given by (\ref{2.7}) naturally
splits into separate flows for the trace-free part and the trace. We
can attempt to prove that the trace-free part and the trace decay
under their respective flows, making the background geometry
linearly stable.  Finally, one can also split a perturbation more
explicitly into a transverse traceless (TT) piece, a trace and the
traceless part of a divergence piece, as $h_{ab} = h^{TT}_{ab} +
\frac{H}{d+2} g_{ab} + \nabla_a Y_b + \nabla_b Y_a -
\frac{\nabla^{c}Y_{c}}{d+2} g_{ab}$. Here $h^{TT}_{ab}$ is
transverse and traceless (TT). We would then have to study the flows
of $h^{TT}_{ab}$, $Y_{a}$ and $H$. Due to the fact that $(\mathbf
Lh^{TT})_{ab}$ is also TT, the flow of $h^{TT}_{ab}$ decouples from
the other flows. In this paper, we are unable to address the full
stability problem as it is computationally difficult. In the next
section, we describe the class of perturbations for which we are
able to obtain analytical stability results.

\section{The flow for a special class of perturbations}
\setcounter{equation}{0}

Let us take the background metric to be of the form  \be ds^2 =
-f(r) dt^2 + g(r) dr^2 + r^2 d{\tilde s}_d^2\ . \label{3.1} \ee Here
$d{\tilde s}_d^2$ is the Riemannian metric on a $d$ dimensional
compact `base manifold' $B$. We will consider the background metric
to be a static solution of the vacuum Einstein equation with a
cosmological constant, \ben R_{ab}= c (d+1) g_{ab}\ ,\label{3.2}
\een where we use Latin letters above for spacetime indices. This
implies that the base manifold is also Einstein, with \ben \tilde
R_{\alpha\beta} &=& \epsilon (d-1) \tilde g_{\alpha\beta},
\label{3.3}\een our convention being that the Greek letters label
the coordinate indices on the base manifold. $\epsilon = \pm 1$ or
$\epsilon = 0$ (in fact, if the spacetime is Ricci-flat, $B$ has to
have positive curvature, so $\epsilon > 0$ \cite{Case}). Further,
$f(r)g(r) = 1$ and $f(r) = (\epsilon - (\alpha /r)^{d-1} -c r^2 )$.

We write the perturbed metric as \be g_{ab} = g_{ab}^{(background)}
+ h_{ab} \label{3.4} \ee where $a$ and $b$ run over $d+2$ indices.
As discussed in the previous section, we can decompose the
perturbation into a traceless part and trace \be h_{ab}={\bf H}_{ab}
+ {1\over d+2} g_{ab} H \label{3.5} \ee
where, $H=g^{ab}h_{ab}$ and $g^{ab} {\bf H}_{ab} = 0$.\\
Then, the flow of the trace $H$ is easily obtained by taking the
trace of (\ref{2.7}) and is \vskip 0.3cm {\bf Flow of trace of an
arbitrary static perturbation:} \noindent \be {\p H \over \p \tau} =
\frac{\alpha'}{2} \left [ ( \Delta H  + 2 c (d+1) H \right ].
\label{3.6} \ee According to our conventions, the Laplacian operator
on the Einstein spacetime (\ref{3.1}) is $\Delta = \nabla^{a}
\nabla_{a}$. In the study of the trace of the perturbation, we will
restrict to perturbations which are static, i.e., time independent.
As mentioned in the introduction, this ensures that the flow
equation for the perturbation is a well-behaved PDE. We can then
study the flow of the trace without further restrictions on the
perturbation. This analysis is presented in sections IV and V.

The next step would be to study the flow of the trace-free part
${\bf H}_{ab}$ --- however, for a general static perturbation, this
equation is difficult to analyze. Recalling that we can write
$h_{ab} = h^{TT}_{ab} + \frac{H}{d+2} g_{ab} + \nabla_a Y_b +
\nabla_b Y_a - \frac{\nabla^{c}Y_{c}}{d+2} g_{ab}$ and that the flow
of $h^{TT}_{ab}$ decouples from the other flows, we can study the
flow of static TT perturbations as a first step. Even this problem
is hard in  all generality. For computational reasons, we focus on
static TT perturbations $h^{TT}_{ab}$ that satisfy $h^{TT}_{ra} =
h^{TT}_{ta} = 0$ for $a$ any spacetime index. These are the class of
`tensor' perturbations on $S^{d}$ --- the term `tensor' perturbation
refers to the fact that the perturbation transforms like a tensor of
rank two on $S^d$ when we do a coordinate transformation on $S^d$.
The full flow (\ref{2.7}) is consistent with this restriction to TT
`tensor' perturbations (and is not consistent with restricting
merely to trace-free perturbations satisfying $h^{TT}_{ra} =
h^{TT}_{ta} = 0$). The Lichnerowicz Laplacian acting on
$h^{TT}_{\alpha \beta}$ can be written for this restricted class of
perturbations and is the same as that obtained by Gibbons and
Hartnoll \cite{GH} in the classical stability analysis of these
black holes, except that the perturbations we consider are
time-independent. Under these restrictions, we can write the flow of
$h^{TT}_{ab}$. \vskip 0.3cm {\bf Flow of TT `tensor' perturbation on
$S^d$ :} \noindent
\begin{eqnarray} {\p h^{TT}_{\alpha \beta} \over \p \tau} &=&
\frac{\alpha'}{2} \left [
 - {1\over r^2} \lb { \tilde \Delta}_L h^{TT} \rb_{\alpha \beta} +
{f(r)} {d^2 \over dr^2} h^{TT}_{\alpha\beta}
 \right. \nonumber \\ && \left.
+ \left ( f'(r) - {(4-d) \over r}f(r) \right )
 {d \over dr} h^{TT}_{\alpha \beta}
+ {4 f(r) \over r^2 } h^{TT}_{\alpha\beta} + 2 c
(d+1)h^{TT}_{\alpha\beta} \right ].~~~~~~~~~~ \label{3.13}
\end{eqnarray}

To study the flow of a general TT perturbation, we will need to also
study the flow of TT perturbations that are `vector' and `scalar'
perturbations on $S^{d}$ (i.e., transforming as rank one and rank 0
tensors respectively, under coordinate transformation in $S^d$). We
discuss how to address the stability problem for a wider class of
perturbations in the last section.

\section{Perturbations of
Schwarzschild Tangherlini black holes and their AdS analogues}
\setcounter{equation}{0} \vskip 0.3cm {\bf Schwarzschild-Tangherlini
black holes: } \vskip 0.3cm In (\ref{3.1}), we now specialize to the
case when $c = 0$. In this case, we must have $\epsilon = 1$. We
will take $B= S^{d}$. $f(r) = (1 - (\alpha /r)^{d-1} )$ and we have
the
Schwarzschild-Tangherlini black holes.\\

{\bf Flow of trace for any static perturbation :} We first study the
flow of the trace of {\em any arbitrary} static perturbation of the
region from the horizon $r = \alpha$ to infinity of these black
holes. We impose Dirichlet conditions on the perturbation at the two
`boundaries', the horizon $r = \alpha$ and $r = \infty$ (we could
instead have chosen Neumann boundary conditions at the horizon ---
the analysis that follows is unaltered by this choice). The flow of
the trace $H$ is given by (\ref{3.6}). Assume the ansatz \be H =
R(r)r^{-d/2} \tilde H (x^a) e^{\frac{\alpha'}{2}\Omega_{T} \tau},
\label {3.14} \ee where \be \tilde \Delta \tilde H = \tilde
\lambda_{T} \tilde H. \label{3.15} \ee $\tilde \Delta = \tilde
g^{\alpha \beta} \nabla_{\alpha}\nabla_{\beta}$ is the Laplacian on
$B$ with respect to the base metric. Then we have \be f(r) R''(r) +
f'(r) R'(r) + \lb \frac{-d^2}{4r^2} f(r) + \frac{d}{2 r^2} f(r) -
\frac{d}{2r} f'(r) + \frac{\tilde \lambda_{T}}{r^2} - \Omega_{T} \rb
R(r) = 0 . \label{3.16} \ee where prime indicate derivative with
respect to $r$. Finally, we define the `tortoise coordinate' $r_{*}$
by $dr_{*} = dr/f(r)$ and so $-\infty < r_{*} < \infty $ when $
\alpha < r < \infty $. We can rewrite (\ref{3.16}) in Schr\"odinger
form as \be - \frac{d^2 R}{dr_{*}^2} - f(r)\lb \frac{-d^2}{4r^2}
f(r) + \frac{d}{2 r^2} f(r) - \frac{d}{2r} f'(r) + \frac{\tilde
\lambda_{T}}{r^2} - \Omega_{T} \rb R = 0. \label{3.17} \ee We are
therefore interested in normalizable (in this case, square
integrable) functions $R(r_{*})$ that correspond to the zero modes
of the Schr\"odinger potential \be V_T (r) = - f(r)\lb
\frac{-d^2}{4r^2} f(r) + \frac{d}{2 r^2} f(r) - \frac{d}{2r} f'(r) +
\frac{\tilde \lambda_{T}}{r^2} - \Omega_{T} \rb . \label{3.18} \ee

Clearly if $R(r_{*}) \rightarrow 0$ as $r_{*} \rightarrow \pm
\infty$ (which is true since we have imposed Dirichlet boundary
conditions on the perturbation) and if $V_T
>0$ for all $-\infty < r_{*} < \infty $, there are no normalizable
zero modes (this can be easily seen by multiplying both sides of
(\ref{3.17}) by $R(r_{*})$ and integrating over the range of
$r_{*}$). We can write $ V_T = V_{1} + \Omega_{T} f(r)$, where \be
V_{1} = - f(r)\lb \frac{-d^2}{4r^2} f(r) + \frac{d}{2 r^2} f(r) -
\frac{d}{2r} f'(r) + \frac{\tilde \lambda_{T}}{r^2} \rb .
\label{3.19} \ee Now $\Omega_{T} f(r) > 0$ for $-\infty < r_{*} <
\infty $, so if $V_{1}
> 0$, no normalizable zero modes are possible. We now explore the
conditions on $\tilde \lambda_{T}$ such that $V_{1}> 0$. Since $0 <
f(r) < 1$ for the range of $r$ we are interested in, \be V_{1}
> 0 \iff \lb \frac{d^2}{4r^2} f(r) - \frac{d}{2 r^2} f(r) +
\frac{d}{2r} f'(r) - \frac{\tilde \lambda_{T}}{r^2} \rb > 0. \label
{3.20} \ee Now, $d \ge 2$ for Schwarzschild-Tangherlini black holes.
In this case, by substituting the explicit form of $f(r)$ in the
above inequality, and also observing that $f'(r)
> 0$, we get \be
\lb \frac{d^2}{4r^2} f(r) - \frac{d}{2 r^2} f(r) + \frac{d}{2r}
f'(r) - \frac{\tilde \lambda_{T}}{r^2} \rb > 0 \iff (\frac{d^2}{4} -
\frac{d}{2} - \tilde \lambda_{T} ) > 0.\label{3.21}\ee

Therefore, there will be no unstable normalizable modes of the trace
of a static perturbation for $\tilde \lambda_{T} <
\frac{d(d-2)}{4}$. $\tilde \lambda_{T}$ is the eigenvalue of the
scalar Laplacian on the base manifold $B$ (according to our
convention, this Laplacian is $\tilde \Delta = \tilde g^{\alpha
\beta} \nabla_{\alpha}\nabla_{\beta}$, acting on smooth functions).
The base manifold for Schwarzschild-Tangherlini black holes is
$S^d$, for which the spectrum of the scalar Laplacian is known. This
spectrum is non-positive (with our conventions) and therefore there
are no unstable normalizable modes of the trace of the perturbation.

{\bf Flow of static TT `tensor' perturbations :} We now examine the
flow of the special class of TT perturbations described in section
III which are rank two `tensor' perturbations on $S^{d}$, ${\bf
H}_{\alpha \beta}$. This is the flow (\ref{3.13}) with $c=0$. We
assume an ansatz of the form $h^{TT}_{\alpha \beta} = \tilde
h^{TT}_{\alpha \beta}(\tilde x) r^{{(4-d)\over 2}} \phi (r)
e^{\frac{\alpha'}{2}\Omega \tau}$, where $\lb {\tilde \Delta}_L
\tilde h^{TT} \rb_{\alpha \beta} = \tilde \lambda \tilde
h^{TT}_{\alpha \beta}$. \footnote{$\tilde h^{TT}_{\alpha \beta}$ are
given in terms of the symmetric tensor spherical harmonics of rank
$2$. It is known that there are no such tensors for $d=2$
\cite{Laplacian}. Therefore while the flow of the trace was relevant
for four dimensional Schwarzschild black holes, these perturbations
are present only for the higher dimensional black holes. } As usual,
perturbations with $\Omega
> 0$ are unstable modes. We then have the following PDE \be - {d
\over dr} \lb f {d\phi \over dr}\rb  + (\frac{d}{2}) {d \over dr}({f
\over r} \phi) + \lb ({f d^2 \over 4r^2 }) + {\tilde \lambda \over
r^2 } - { 2 f'(r) \over r} - {(2d-2)f \over r^2 } \rb \phi = -
\Omega \phi. \label{3.23} \ee

Rewriting the above equation in Schr\"odinger form using the
tortoise coordinate $r_{*}$, we have \begin{eqnarray} {-d^2 \over
dr_{*}^{2}} \phi + \bar V (r) \phi = 0~~~~~~~~~~~~~, \label{3.24a} \\
\bar V (r) = f(r) \lb {\tilde \lambda \over r^2 } + {(d-4)f'(r)
\over 2 r} + {(d^2 -10d + 8) \over 4} {f \over r^2 } + \Omega \rb .
\label{3.24}
\end{eqnarray}

This equation, and the flow of the TT modes is similar to that in
\cite{GH} (with some important differences : $\Omega > 0$, which
labels the unstable mode now appears in the potential, and we are
interested in {\em zero} eigenvalues of this potential). As before
in the case of the flow of the trace, $\Omega f(r) > 0$ for $-\infty
< r_{*} < \infty $, so if \be \bar V_{1} = f(r) \lb {\tilde \lambda
\over r^2 } + {(d-4)f'(r) \over 2 r} + {(d^2 -10d + 8) \over 4} {f
\over r^2 } \rb
> 0, \label{3.25} \ee no normalizable eigenfunctions with zero eigenvalue
are possible for the
potential $\tilde V$. From the explicit form of $f(r)$, it is clear
that \be \bar V_{1} (r) > 0 \iff \tilde \lambda > - {(d^2 -10d + 8)
\over 4}. \label{3.26} \ee Recall that $\tilde \lambda$ is the
eigenvalue of the Lichnerowicz operator on the base manifold, $\lb
{\tilde \Delta}_L \tilde h^{TT} \rb_{\alpha \beta} = \tilde \lambda
\tilde h^{TT}_{\alpha \beta}$. For the Schwarzschild-Tangherlini
black holes, the base manifold is $S^d$, and one can write down the
precise form of the Lichnerowicz operator. $\lb {\tilde \Delta}_L
\tilde h^{TT} \rb_{\alpha \beta} = - ( \tilde \Delta \tilde
h^{TT})_{\alpha \beta} + 2 d \tilde h^{TT}_{\alpha \beta}$, and the
spectrum of the Laplacian $\tilde \Delta $ on symmetric two-tensors
$S^d$ is non-positive. In fact, the space of symmetric two-tensors
on $S^d$ is spanned by a canonical set of symmetric, transverse,
trace-free spherical harmonics obeying \be (\tilde \Delta \tilde
h)_{\alpha \beta} = - [k(k+d-1) - 2]; \label{3.27} \ee for integers
$k \ge 2$ (see p. 30, \cite{GH} and also \cite{Laplacian} for the
spectrum of the Laplacian on symmetric two-tensors in $S^d$). Thus
$\tilde \lambda \ge 2d$.

From our analysis above, it is clear that unstable normalizable
modes of this TT perturbation are only possible when the potential
$\tilde V$ is not positive. A necessary condition therefore is that
$\tilde \lambda \le - {(d^2 -10d + 8) \over 4}$. However, $\tilde
\lambda \ge 2d$, and it is easy to see that this condition can never
be satisfied. Therefore the Schwarzschild-Tangherlini black holes
are stable under the class of TT perturbations we have considered.
\vskip 0.3cm {\bf AdS-Schwarzschild-Tangherlini black holes:} \vskip
0.3cm We set the cosmological constant $c = -1 / L^2$. Also let $B =
S^d$. Then $f(r)
= (1 - (\alpha /r)^{d-1} + r^{2}/L^{2})$.\\

{\bf Flow of trace for any static perturbation :} Considering the
flow of the trace $H$ of an arbitrary static perturbation
(\ref{3.6}) as before, and assuming the same ansatz (\ref{3.15}),
the equation for $R(r)$ in tortoise coordinates is \be - \frac{d^2
R}{dr_{*}^2} - f(r)\lb \frac{-d^2}{4r^2} f(r) + \frac{d}{2 r^2} f(r)
- \frac{d}{2r} f'(r) + \frac{\tilde \lambda_{T}}{r^2} - \Omega_{T} -
(2 / L^2) (d+1) \rb R = 0.\label{3.28} \ee There are no physically
reasonable solutions when the potential
\\$V_T = - f(r)\lb \frac{-d^2}{4r^2} f(r) + \frac{d}{2 r^2} f(r) -
\frac{d}{2r} f'(r) + \frac{\tilde \lambda_{T}}{r^2} \rb $ is
positive (either the perturbation or its first derivative is not
normalizable). As in the previous analysis, by putting the explicit
form of $f(r)$ above, we deduce that a necessary condition for $V_T$
to be negative is that $( \frac{d^2}{4} - \frac{d}{2} - \tilde
\lambda_{T} ) < 0$, which is the same condition as in the zero
cosmological constant case. Thus a necessary condition for the
existence of the normalizable unstable modes of the trace is that
$\tilde \lambda_{T} > \frac{d(d-2)}{4}$, which is not fulfilled by
the spectrum of the Laplacian on $S^d$, and such modes do not exist.

{\bf Flow of static TT `tensor' perturbations :} We can also repeat
the analysis for the flow of the TT perturbations that are rank two
tensor perturbations on $S^d$, given by (\ref{3.13}). We assume the
ansatz (\ref{3.23}) as in the zero cosmological constant case. We
get the equation in Schr\"odinger form to be (\ref{3.24a}) with \be
\bar V (r) = f(r) \lb {\tilde \lambda \over r^2 } + {(d-4)f'(r)
\over 2 r} + {(d^2 -10d + 8) \over 4} {f \over r^2 } + (2 / L^2 )
(d+1)+ \Omega_T \rb. \label{3.29} \ee As before, a necessary
condition for the potential $\bar V $ to be negative is the same as
the zero cosmological constant case; $\tilde \lambda \le - {(d^2
-10d + 8) \over 4}$. This is never possible for the base manifold
being $S^d$, and therefore the AdS-Schwarzschild-Tangherlini black
holes are also stable under the class of perturbations considered.
We could attempt a similar analysis for de Sitter black holes.
However, in this case, the perturbations are confined to the region
between the black hole and cosmological horizons. The conditions
under which the potential ((\ref{3.29}) with $L^2$ replaced by
$-L^2$) is positive in this case are not easy to read off, and will
depend on the relative magnitudes of the two horizon radii.

\section{A more rigorous argument}
\setcounter{equation}{0}

In this section, we present a more rigorous discussion of the
stability of the black holes considered in the previous section,
either for the flow of the trace of any static perturbation, or the
flow of the static, TT perturbations that are tensor perturbations
on $S^d$. The first step in our study was the choice of ansatz
--- of the form (\ref{3.14}) for the flow of the trace (\ref{3.6});
and of the form (\ref{3.23}) for the flow of the traceless part
(\ref{3.13}). The question that naturally arises is : how general is
this ansatz? Clearly there is no loss of generality in expressing
the dependence of the perturbation on angular coordinates on the
base $S^d$ in terms of suitable scalar or tensor spherical harmonics
as we have done. However, our ansatz for the $\tau$ dependence of
the perturbation involves the assumption that every unstable
perturbation that is a solution of the flow (either of the trace or
TT tensor part) can be expressed as a superposition of solutions
with $\tau$ dependence of the form $e^{\Omega \tau}$ where $\Omega >
0$ is real. A further assumption we are implicitly making in such a
stability analysis is that if we cannot find normalizable modes
satisfying our ansatz, then there can be no perturbations of compact
support, for example, that are growing in $\tau$. This need not be
true, as we could conceive of perturbations of compact support
constructed as a linear superposition of the unnormalizable modes
(these would consequently grow in $\tau$). Generically, operators
such as the Laplacian or the Lichnerowicz Laplacian have a spectrum
with a continuous component on noncompact manifolds, and the
corresponding eigentensors are not normalizable. Absence of such
unstable normalizable modes does not guarantee stability of the
spacetime under initially well-behaved perturbations. We therefore
present a more rigorous stability result below to resolve this
issue. In the context of classical stability analysis of black
holes, such a rigorous argument was given by Wald \cite{wald}. Our
argument bears some similarities to it --- however, we analyze a
(degenerate) parabolic PDE, not a hyperbolic PDE as in the classical
stability analysis, and thus there are differences in the proof.

After separation of angular variables, the flow of either the trace
or TT tensor perturbation is given in terms of a function
$\Phi(r_{*}, \tau)$ ( where the tortoise coordinate $r_{*}$ is
defined by $dr_{*} = dr/f $) by an equation of the form: \be
{\partial \over
\partial \tau} \Phi (r_{*}, \tau) = - \mathcal{L}\Phi(r_{*}, \tau),
\label{5.1}\ee where \be \mathcal{L} = {1 \over f} \lb - {\partial^2
\over \partial r_{*}^2} + V \rb . \label{5.2} \ee $V$ is a
`potential' of the form $V_1$ or $\bar V_1$ we saw in the previous
section. In the $r$ coordinate, \be \mathcal{L} = - {\partial \over
\partial r}\lb f {\partial \over
\partial r}\rb + {V \over f}. \label{5.2a} \ee Then, we are interested in the situation when $V
> 0$. We saw that both for the flow of the trace and TT tensor
perturbation of the (AdS) Schwarzschild-Tangherlini black holes,
this is always true for the potential. We will assume $\Phi(r_{*},
\tau)$ to be a smooth ($C^{\infty}$) function of $r_{*}$ (so that
the perturbed geometry is smooth). We will also assume that it is of
compact support in $r_{*}$. In fact, our analysis applies to a wider
class of perturbations. Specifically, we only need to impose a
Dirichlet (or Neumann) boundary condition on the perturbation at
$r_{*} = - \infty$ (treating the horizon as a boundary) and suitable
decay conditions as $r_{*} \rightarrow \infty$ so that $\Phi$ and
its derivatives up to order two go to zero in this limit. This
ensures that boundary terms generated while integrating by parts in
the following analysis vanish.

The question we address is whether there exist growing solutions
$\Phi(r_{*},\tau)$ to (\ref{5.1}) when $V > 0$. We first define the
following `energies':
\begin{eqnarray} E_{0} = \int_{\alpha}^{\infty} \Phi(r, \tau)^2 dr ; \nonumber \\
E_{1} = \int_{\alpha}^{\infty} (\mathcal{L}\Phi(r, \tau))^2 dr.
\label{5.3} \end{eqnarray}

Next, we observe that:\begin{eqnarray} {d \over d\tau} E_{0} &=& -
2 \int_{\alpha}^{\infty} \Phi (\mathcal{L}\Phi) ~dr ;\nonumber \\
&=& 2 \int_{\alpha}^{\infty} \Phi {\partial \over \partial r}\lb f
{\partial \over \partial r}\rb \Phi ~dr - 2 \int_{\alpha}^{\infty}
{V \over f} \Phi^2 ~dr. \label{5.4} \end{eqnarray} Note that for the
black hole potentials, ${V \over f}$ is finite as $r \rightarrow
\alpha$.  Integrating the first integral on the right by parts, we
get no boundary terms, since the perturbation is of compact support.
Therefore, since $f \geq 0$ and $V
> 0$,
\begin{eqnarray} {d \over d\tau} E_{0} &=& - 2 \int_{\alpha}^{\infty}
f \lb {\partial \Phi \over \partial r} \rb^2 ~dr - 2
\int_{\alpha}^{\infty} {V \over f} \Phi^2 ~dr \leq 0. \label{5.5}
\end{eqnarray}
By replacing $\Phi$ with $\mathcal{L}\Phi$ in the above argument, we
see that $ {d \over d\tau} E_{1} \leq 0$ as well. Therefore
$E_{0}(\tau)$ and $E_{1}(\tau)$ are bounded from above by their
initial values $E_{0}(0)$ and $E_{1}(0)$ which we assume are finite.
By the Cauchy-Schwarz inequality, \be \left | \int_{\alpha}^{\infty}
\Phi \mathcal{L}\Phi ~dr \right | \leq \sqrt{E_{0}(\tau)
E_{1}(\tau)} \leq \sqrt{E_{0}(0) E_{1}(0)} \label{5.6} \ee

 Now define the following integrals:
\begin{eqnarray} I_{0} = \int_{r_{*} = - \infty}^{r_{*} = \infty} \Phi (r_{*},
\tau)^2 ~dr_{*} ; \nonumber \\ I_{1} = \int_{r_{*} = -
\infty}^{r_{*} = \infty} ({\partial \Phi (r_{*}, \tau) \over
\partial r_{*}})^2 ~dr_{*}. \label{5.7}
\end{eqnarray}

We note that
\begin{eqnarray}
I_{1} = \int_{- \infty}^{\infty} ({\partial \Phi (r_{*}, \tau) \over
\partial r_{*}})^2 ~dr_{*} \nonumber \\ = \int_{- \infty}^{\infty} f \Phi
\mathcal{L}\Phi ~dr_{*} - \int_{- \infty}^{\infty}V \Phi^2
~dr_{*};\nonumber \\
I_{1} \leq \int_{- \infty}^{\infty} f \Phi \mathcal{L}\Phi ~dr_{*} =
\int_{\alpha}^{\infty} \Phi \mathcal{L}\Phi ~dr,\label{5.8}
\end{eqnarray}
where we did integration by parts and discarded boundary terms as
the perturbation has compact support. Thus, from (\ref{5.6}) and
(\ref{5.8}), we conclude that $I_{1}(\tau)$ is bounded from above by
a $\tau$ independent constant $C_{1}$ depending on the initial
values of the energies. Our analysis so far applies both to
Schwarzschild-Tangherlini and AdS-Schwarzschild-Tangherlini black
holes, provided the integrals $E_0$, $E_{1}$, $I_{0}$ and $I_{1}$
are finite at initial $\tau$. Now there is a slight departure in the
analysis for the two types of black holes in the next step.

We now wish to prove that $I_{0}$ is also bounded from above by a
$\tau$ independent constant $C_{0}$. \vskip 0.3cm {\bf For
Schwarzschild-Tangherlini black holes:}  \vskip 0.3cm From
(\ref{5.5}), we conclude that for $\tau \ge 0$
\begin{eqnarray}
E_{0} (0) - E_{0}(\tau) = \int_{- \infty}^{\infty} f \lb \Phi(r_{*},
0)^2 - \Phi(r_{*}, \tau)^2 \rb ~dr_{*} \geq 0. \label{5.9}
\end{eqnarray}
Since $0\leq f \leq 1$, we note that \be \Phi(r_{*}, 0)^2 -
\Phi(r_{*}, \tau)^2 \geq f \lb \Phi(r_{*}, 0)^2 - \Phi(r_{*},
\tau)^2 \rb \label{5.10} \ee for all $r_{*}$. Therefore
\begin{eqnarray}
\int_{- \infty}^{\infty} \lb \Phi(r_{*}, 0)^2 - \Phi(r_{*}, \tau)^2
\rb dr_{*} &&\geq  \int_{- \infty}^{\infty} f \lb \Phi(r_{*}, 0)^2 -
\Phi(r_{*}, \tau)^2 \rb dr_{*} \nonumber \\
&&\geq 0. \label{5.11}
\end{eqnarray}
Thus $I_{0}(\tau)$ is bounded from above by its initial value which
we denote $C_{0}$. \newpage {\bf For AdS-Schwarzschild-Tangherlini
black holes:} \vskip 0.3cm From (\ref{5.5}) we conclude that for
$\tau \ge 0$, \begin{eqnarray} && E_{0} (0) - E_{0}(\tau) = \int_{-
\infty}^{\infty}(1 - (\alpha /r)^{d-1} + r^{2}/L^{2}) \lb
\Phi(r_{*}, 0)^2 - \Phi(r_{*}, \tau)^2 \rb ~dr_{*} \geq 0
\nonumber \\
&& \implies \int_{- \infty}^{\infty}(1 + r^{2}/L^{2} )  \lb
\Phi(r_{*}, 0)^2 - \Phi(r_{*}, \tau)^2 \rb ~dr_{*} \geq 0,
\label{5.9a}
\end{eqnarray}
as $(1 - (\alpha /r)^{d-1} + r^{2}/L^{2}) \leq (1 + r^{2}/L^{2})$
for all $r$ in the range of interest.

 Therefore, we have the following inequality:
\begin{eqnarray}
\int_{- \infty}^{\infty} \Phi(r_{*}, \tau)^2 ~dr_{*} &&\leq \int_{-
\infty}^{\infty} (1 + r^{2}/L^{2} ) \Phi(r_{*}, \tau)^2 ~dr_{*}
\nonumber
\\
&& \leq \int_{- \infty}^{\infty}(1 + r^{2}/L^{2} ) \Phi(r_{*}, 0)^2
~dr_{*}. \label{5.10a} \end{eqnarray}

Assume that the integral $\int_{- \infty}^{\infty}(1 + r^{2}/L^{2} )
\Phi(r_{*}, 0)^2 ~dr_{*} = C_{0}^{AdS}$ is finite. Then $I_{0}$ is
bounded from above by the $\tau$ independent constant $C_{0}^{AdS}$.
\vskip 0.3cm The rest of the analysis applies to both classes of
black holes.
 Now we expand $\Phi
(r_{*}, \tau)$ in terms of its Fourier modes as
\begin{eqnarray} \Phi(r_{*}, \tau) &=& {1 \over \sqrt{2\pi }} \int_{- \infty}^{\infty}
e^{ikr_{*}} \hat \Phi (k, \tau) ~dk ; \nonumber \\
&&= {1 \over \sqrt{2\pi }} \int_{- \infty}^{\infty}  (1 +
k^2)^{-1/2} e^{ikr_{*}}(1+ k^2)^{1/2} \hat \Phi (k, \tau) ~dk.
\label{5.12}
\end{eqnarray}

Finally, using the Cauchy-Schwarz inequality, we obtain
\begin{eqnarray}
&& \left |\Phi(r_{*}, \tau)\right |^2 =\left ( {1 \over \sqrt{2\pi
}} \int_{- \infty}^{\infty}  [(1 + k^2)^{-1/2}e^{ikr_{*}}] [(1+
k^2)^{1/2} \hat \Phi (k, \tau)] ~dk \right )^2 ; \nonumber \\ &&
\leq {1 \over 2\pi} \lb \int_{- \infty}^{\infty}(1 + k^2 )^{-1}e^{
2ikr_{*}} dk \rb \lb \int_{- \infty}^{\infty} (1 + k^2 ) \left |\hat
\Phi(k, \tau)\right |^2  dk \rb. \label{5.13}
\end{eqnarray}

Now, using elementary contour integration techniques, \be {1 \over 2
\pi} \int_{- \infty}^{\infty}(1 + k^2 )^{-1}e^{ 2ikr_{*}} dk = {1
\over 2} ~e^{- 2|r_{*}|}.\label{5.14} \ee

Using the Plancherel theorem, \be \int_{- \infty}^{\infty} (1 + k^2
)~ \left |\hat \Phi(k, \tau)\right |^2 dk = \int_{- \infty}^{\infty}
|\Phi (r_{*}, \tau)|^2 ~dr_{*} + \int_{- \infty}^{\infty} \left |
{\partial \Phi (r_{*}, \tau) \over
\partial r_{*}}\right |^2 ~dr_{*}. \label{5.15} \ee

Thus, we have the following inequality (a simple case of a {\em
Sobolev inequality}, where the right-hand side is the square of a
Sobolev norm):
\begin{eqnarray}
\left |\Phi(r_{*}, \tau)\right |^2 \leq \int_{- \infty}^{\infty}
|\Phi (r_{*}, \tau)|^2 ~dr_{*} + \int_{- \infty}^{\infty} \left |
{\partial \Phi (r_{*}, \tau) \over
\partial r_{*}}\right |^2 ~dr_{*}. \label{5.16} \end{eqnarray}

We have already seen that the integrals $I_{0}(\tau)$ and
$I_{1}(\tau)$ are bounded from above by $\tau$-independent constants
$C_{0}$ and $C_{1}$ respectively (we replace $C_{0}$ with
$C_{0}^{AdS}$ below in the AdS case). It follows that \be \left
|\Phi(r_{*}, \tau)\right |^2 \leq C_{0} + C_{1}.\label{5.17} \ee

Therefore, we have shown that the pointwise norm of the perturbation
$\Phi(r_{*}, \tau)$ stays bounded along the linearized flow. While
we have not proved that this norm goes to zero, this clearly
clarifies some of the issues we raised about the choice of ansatz
(\ref{3.14}) or (\ref{3.23}) for the trace or the TT perturbation,
respectively. There are no trace modes or TT `tensor' perturbations
of compact support growing in $\tau$ for either the
Schwarzschild-Tangherlini or AdS-Schwarzschild-Tangherlini black
holes. Both in Wald's paper on the classical stability of the
Schwarzschild black hole \cite{wald} and in the proof of linear
stability of $\mathbf{H^n}$ under Ricci flow, it is possible to also
bound the higher derivatives of the perturbation under the flow. We
are unable to use techniques similar to those in \cite{suneeta} to
bound higher derivatives as the operator $\mathcal{L}$ given by
(\ref{5.2}) is not self-adjoint with respect to the measure
$dr_{*}$.

\section{Summary and discussion}
\setcounter{equation}{0}

The results presented in this paper are the beginning of a program
to study the linear stability of Schwarzschild-Tangherlini black
holes (and their AdS counterparts) under Ricci flow. As discussed in
the introduction, there are diverse motivations from physics for
such a study. Such stability results also offer insights on Ricci
flow on noncompact manifolds, which is not as well-understood as
Ricci flow on compact manifolds. We briefly summarize the results of
this paper:\\

We study the evolution of static perturbations of the spacetime
outside the horizon of the (AdS) Schwarzschild-Tangherlini
spacetimes under a linearized Ricci flow (or a flow related to it by
rescalings). \\ \noindent (i) We are able to show that there is no
instability under the flow of the trace of an arbitrary static
perturbation. This is done in two steps: We assume a separation of
variables ansatz in section IV and show that the flow equation for
the trace has no unstable normalizable modes. In section V, we go
beyond such a specific choice of anzatz and show that the pointwise
norm of a solution of compact support in $r_{*}$ stays bounded under
the linearized flow of the trace.\\ \noindent
(ii) For static TT
perturbations that obey $h^{TT}_{ra} = h^{TT}_{ta} = 0$, where $a$
is any spacetime index (i.e., perturbations that behave as rank two
tensors on $S^d$), we show that there is no instability under the
flow. This is done by adopting a specific ansatz in section IV, and
showing in section V (as for the trace) that the pointwise norm of a
perturbation of
compact support in $r_{*}$ stays bounded under the flow.\\

We now discuss how to widen our study to a more general class of
static perturbations. The strategy is to first split the
perturbation explicitly into a TT part, a part proportional to the
trace, and the traceless part of a divergence as $h_{ab} =
h^{TT}_{ab} + \frac{H}{d+2} g_{ab} + \nabla_a Y_b + \nabla_b Y_a -
\frac{\nabla^{c}Y_{c}}{d+2} g_{ab}$. Then, we consider its evolution
under either (\ref{2.5}) or (\ref{2.7}) and attempt to decouple the
flows of the various parts by choosing $V_a$ appropriately, as for
example, in (\ref{2.5}). The flow of $h^{TT}_{ab}$ decouples and can
be studied separately. The analysis of this flow is simplified by a
result of Kodama and Sasaki (see p. 139 in \cite{ks}). The result
implies that any covariant linear differential equation on the
spacetime that is at most second order (like the flow of
$h^{TT}_{ab}$) is decomposed into equations for perturbations
$h^{TT}_{ab}$ that behave as scalar, vector and rank two tensor on
$S^d$ respectively (this is also true of the classical stability
analysis in \cite{ishibashi}). Thus we can then analyze the flow of
each type of TT perturbation separately. We have analyzed the rank
two tensor type in this paper. We hope to analyze the scalar and
vector type TT perturbations in future work. We have already
concluded that there is no instability in the flow of the trace. We
hope that a systematic analysis as outlined will lead to a better
understanding of the nature of the unstable modes of the (AdS)
Schwarzschild-Tangherlini spacetimes under Ricci flow or in quantum
gravity.

\section{Acknowledgements}
VS thanks  Vincent Moncrief for a useful discussion on gauge-fixing.
VS also thanks the organizers of the Pisa workshop on Geometric
flows in mathematics and theoretical physics where a preliminary
version of this work was presented, and the
Albert-Einstein-Institut, Golm for hospitality where a part of this
work was carried out. This work is supported by funds from the
Natural Sciences and Engineering Research Council of Canada.


\begin{thebibliography}{100}
\bibitem{sbhstable} T Regge, J Wheeler, Phys Rev 108 (1957) 1063;
CV Vishveshvara, Phys Rev D1 (1970) 2870; R Price, Phys Rev D5
(1972) 2419; V Moncrief, Ann Phys 88 (1973) 323; RM Wald, J Math
Phys {\bf 20} (1979) 1056.
\bibitem{GH} G Gibbons, S Hartnoll, Phys Rev D66:064024 (2002).
\bibitem{ishibashi} A Ishibashi and H Kodama, Prog Theor Phys 110 (2003)
901.
\bibitem{GPY} DJ Gross, MJ Perry, LG Yaffe, Phys Rev D25 (1982) 330.
\bibitem{reall} HS Reall, Phys Rev D64 (2001) 044005.
\bibitem{GL} R Gregory, R Laflamme, Nucl Phys B428 (1994) 399.
\bibitem{prestidge} T Prestidge, Phys Rev D61 (2000) 084002; SS
Gubser, Class Quant Grav 19 (2002) 4825; B Kol, E Sorkin, Class
Quant Grav 21 (2004) 4793; B Kol, Phys Rev D77 (2008) 044039 and
references therein.
\bibitem{hw} M Headrick, T Wiseman, Class Quant Grav 23 (2006) 6683.
\bibitem{hwk} M Headrick, S Kitchen, T Wiseman, arXiv:0905.1822;
see also G Holzegel, T Schmelzer, C Warnick, Class
Quant Grav 24 (2007) 6201.
\bibitem{doran} C Doran, M Headrick, CP Herzog, J Kantor, T Wiseman,
Commun Math Phys 282 (2008) 357.
\bibitem{tian} G Tian, X Zhu, J Amer Math Soc 20 (2007) 675, S
08964-0347(06)00552-2.
\bibitem{HMT} M Headrick, S Minwalla, T Takayanagi, Class Quant Grav 21 (2004)
S1539.
\bibitem{APS} A Adams, J Polchinski, E Silverstein, JHEP 0110 (2001) 029;
Y Okawa, B Zwiebach, JHEP 0403 (2004) 056.
\bibitem{cao} H-D Cao, J Diff Geom 45 (1997) 257.
\bibitem{GHMS} M Gutperle, M Headrick, S Minwalla, V Schomerus, JHEP
0301 (2003) 073.
\bibitem{suneeta} V Suneeta, Class Quant Grav 26 (2009) 035023.
\bibitem{wald} RM Wald, J Math Phys 20 (1979) 1056.
\bibitem{chow} B Chow, P Lu, L Ni, {\em Hamilton's Ricci Flow},
Graduate Studies in Mathematics, Volume 77, American Mathematical
Society Science Press (2006).
\bibitem{topping} P Topping, {\em Lectures on the Ricci Flow},
London Mathematical Society Lecture note series 325, Cambridge
University Press (2006).
\bibitem{rfstability} R Ye, Trans Amer Math Soc 338 (1993) 871;
C Guenther, J Isenberg, D Knopf, Comm Anal Geom 10 (2002) no.4, 741;
Int. Math. Res. Not. (2006), Article ID 96253, doi:
10.1155/IMRN/2006/96253; H-D Cao, R Hamilton, T Ilmanen, arXiv:
math/0404165; N Sesum, arXiv: math/0410062; X Dai, X Wang, G Wei,
arXiv: math/0504527; O Schnuerer, F Schulze, M Simon, Comm Anal Geom
16 (2008) 127; D Knopf, A Young, Proc Amer Math Soc,137 (2009)699, M
Zhu, arXiv: 0901.2942.
\bibitem{Case} JS Case, arXiv:0902.2226v2.
\bibitem{Laplacian} MA Rubin, CR Ordonez, J Math Phys 25
(1984) 2888; A Chodos, E Myers, Ann Phys 156 (1984) 412; A Higuchi,
J Math Phys 28 (1987) 1553.
\bibitem{ks} H Kodama, M Sasaki, Prog Theor Phys Suppl 78 (1984) 1.

\end{thebibliography}
\end{document}